\documentclass[12pt]{article}
\usepackage{epsfig}
\def\be{\begin{eqnarray}}
\def\ee{\end{eqnarray}}
\def\bq{\begin{equation}}
\def\eq{\end{equation}}
\def\ben{\begin{enumerate}}\def\een{\end{enumerate}}

\def\roughly#1{\mathrel{\raise.3ex\hbox{$#1$\kern-.75em%
\lower1ex\hbox{$\sim$}}}}

\begin{document}
\begin{titlepage}

\hfill FTUV  07-0518


 \vspace{1.5cm}
\begin{center}
\ \\
{\bf\LARGE Hidden Dirac Monopoles}
\\
\vspace{0.7cm} {\bf\large Vicente Vento} \vskip 0.7cm

{\it  Departamento de F\'{\i}sica Te\'orica and Instituto de
F\'{\i}sica Corpuscular}

{\it Universidad de Valencia - Consejo Superior de Investigaciones
Cient\'{\i}ficas}

{\it 46100 Burjassot (Val\`encia), Spain, }

{\small Email: Vicente.Vento@uv.es}

\end{center}
\vskip 1cm \centerline{\bf Abstract}

Dirac showed that the existence of one magnetic pole in the universe
could offer an explanation of the discrete nature of the electric
charge. Magnetic poles appear naturally in most grand unified
theories. Their discovery would be of greatest importance for
particle physics and cosmology. The intense experimental search
carried thus far has not met with success. I proposed a universe
with magnetic poles which are not observed free because they hide in
deeply bound monopole--anti-monopole states named monopolium. I
discuss the realization of this proposal and its consistency with
known cosmological features. I furthermore analyze its implications
and the experimental signatures that confirm the scenario.

\vspace{2cm}

\noindent Pacs: 14.80.Hv, 95.30.Cq, 98.70.-f, 98.80.-k


\end{titlepage}

\section{Introduction}

The theoretical justification for the existence of classical
magnetic poles, hereafter called monopoles, is that they add
symmetry to Maxwell's equations and explain charge quantization
\cite{dirac1}. Dirac showed that the mere existence of a monopole
in the universe could offer an explanation of the discrete nature
of the electric charge. His analysis leads to the so called Dirac
Quantization Condition (DQC),

\bq \frac{e g}{\hbar c} = \frac{N}{2} \;, \mbox{  N = 1,2,...}\; ,
\eq

\noindent where $e$ is the electron charge and $g$ the monopole
charge \cite{dirac1}. Note that if quarks were asymptotic states
the minimum monopole charge would be three times larger.

The origin of monopoles, and therefore their properties, is diverse.
In Dirac's formulation monopoles are assumed to exist as point-like
particles and quantum mechanical consistency conditions lead to
Eq.(1), establishing the value of their magnetic charge. However,
their mass, $m$, is a parameter of the theory, limited only by
classical reasonings to be $m > 2 $ GeV \cite{book}. In non-Abelian
gauge theories monopoles arise as topologically stable solutions
through spontaneous breaking via the Kibble mechanism \cite{kibble}.
They are allowed by most Grand Unified Theory (GUT) models, have
finite size and come out extremely massive $m > 10^{16}$ GeV.
Furthermore, there are also models based on other mechanisms with
masses between those two extremes \cite{book,giacomelli,rujula}.

The discovery of monopoles would be of greatest importance not only
for particle physics but for cosmology as well. Therefore monopoles
and their experimental detection have been a subject of much study
since many believe in Dirac's statement\cite{dirac1}\\

{\sl "...one would be surprised if Nature had made no use of
it [the monopole]."}\\

At present, despite intense experimental search, there is no
evidence of their existence
\cite{book,giacomelli,review,experiment,mulhearn}. This state of
affairs has led me to investigate a possible mechanism by which
monopoles could exist and still be undetectable by present
experiments.

Although monopoles symmetrize   Maxwell's equations in form there is
a numerical asymmetry arising from the DQC, namely that the basic
magnetic charge is much larger than the smallest electric charge.
This led Dirac himself in his 1931 paper
\cite{dirac1} to state,\\

{\sl "... the attractive force between two one quantum poles of
opposite sign is $(\frac{137}{2})^2 \approx 4692\frac{1}{4}$ time
that between the electron and the proton. This very large force
may perhaps account for why the monopoles have never been
separated."} \\

This statement by Dirac has motivated the present investigation. I
propose a scenario where monopoles exist but are hidden from our
direct observation because today they appear forming deep
monopole-anti-monopole bound states. I introduce in the next
paragraph my proposal which I elaborate in detail in the next
sections together with its experimental connotations.

At some early stage in the expansion of the Universe, monopoles and
their antiparticles were created. At a later time the dynamics was
such that most of the poles paired up to form monopole-anti-monopole
bound states called monopolia. This happened because it is easier
for the monopoles to interact with each other than with the charged
particles in the hot plasma, thus as a consequence also no
significant friction force arises. Therefore, the lifetime of the
primordial monopolium is solely governed by cascading to the lower
bound states where the poles finally annihilate. Moreover, the
lifetime of monopolium is sufficiently long to allow primordial
monopolia to exist even today in measurable abundances. Thus today,
almost all of the existing monopoles, appear confined in deeply
bound states \cite{zeldovich}. However, the present investigation
would be irrelevant if no proof of the existence of monopolium could
be found. I analyze signatures of their present existence and of
their formation period.

\section{Hidden monopoles}

 I envisage a scenario, in which monopoles are not
observable as free states at present, which is realized by means of
a few assumptions, that satisfies all phenomenological restrictions
and leads to new observations which can sustain it. If my scenario
is confirmed experimentally it will strongly restrict the way
cosmological models deal with monopoles.

At some early stage in the expansion of the universe monopoles and
their antiparticles were created by a mechanism which is free from
the standard monopole problem \cite{kolb}. No precise mechanism for
their creation is advocated, therefore the mass is not fixed and is
left as a parameter to be fitted by consistency requirements.
Moreover, monopoles and anti-monopoles existed in the universe, at
that time, at the level of abundance compatible with known
phenomenological and experimental upper bounds
\cite{ahlen,parker,adams}. These are the same starting assumptions
of all similar treatments \cite{hill,sigl,blanco}. I next depart
from them by assuming that during nucleosynthesis, most of the
monopoles and anti-monopoles bind in pairs, due to the strong
magnetic forces, to form monopolium. This scheme is realized
physically by imposing that: i) the capture radius of the poles,
$r_{capture}$, and the mean free path for charge particle collision
in the hot plasma, $\lambda$, satisfy,

\bq r_{capture} << \lambda;\label{inequality} \eq
ii) a consistent monopolium formation scenario. Let me discuss in
this section the first assumption which leads to the determination
of some of the monopole and monopolium properties and leave for the
next section the discussion on monopolium formation.

The capture radius, $r_{capture}$, is given by
\cite{book,zeldovich},

\bq r_{capture} \sim \frac{g^2}{k T} \; .\eq

The mean free path $\lambda$ is given by

\bq \lambda \sim \frac{1}{\sigma \rho_{ch}}\; , \eq

\noindent where $\sigma$ is the cross section for the scattering of
the monopole with charged particles \cite{book,zeldovich}

\bq \sigma \sim 2 \, \frac{m c^2}{k T} \;\; {\mbox nanobarns}\; ,
\eq

\noindent $\rho_{ch}$ is the density of charged particles and $m$ is
the mass of the monopole.

I describe the monopoles from the monopolium formation era up to the
present days by point like Dirac monopoles. It is reasonable to do
so since the discussion is largely independent of the detailed
structure of the monopoles because it depends only on global
properties, i.e., magnetic charge, mass and cosmological abundances.

In the rest of the paper I proceed to show that my assumptions lead
to a picture which is consistent with present data.

I describe monopolium as a Bohr atom, with reduced mass $m/2$ and a
strong magnetic, instead of a weak electric, coupling. Its binding
energy is

\bq E \sim \left(\frac{1}{8 \alpha}\right)^2 \;\frac{m c^2}{n^2}\;
,\eq

\noindent where $\alpha = \frac{1}{137}$ is the fine structure
constant of $QED$ and $n$ the principal quantum number. This
equation and those that follow are to be considered only for large
principal quantum number ($n > 50$). For low values of $n$ the
annihilation mechanism becomes dominant.

The approximate size of the system is given by

\bq r \; \sim \; <r>_{n,0}\; \sim\; \frac{12 \hbar}{m c}\; \alpha
\; n^2 \; . \eq

\noindent To calculate the mean life I distinguish two processes, i)
the cascading process, dominated by dipole radiation \cite{hill},
which I apply as

\bq \tau_{dipole} \sim \frac{ 2 m^2 c a_n^3}{\hbar^2} \sim 2\;
(12)^3 \;\frac{\hbar}{mc^2}\; \alpha ^5 \;n_i^6 \; ,\eq

\noindent where $n_i$ is the principal quantum number associated
with the initial bound state which will be very large $n_i \sim
10^9$; ii) the annihilation process, which due to the magnitude of
$g$ is highly non perturbative and which I next estimate. Looking at
the two photon decay process I see \cite{dirac2,wheeler}

\bq \tau_{annihilation} < < \tau_{2 \gamma} \sim 2\; (4)^5
\frac{\hbar}{m c^2} \;\alpha^5 \; n_f^3 \; ,\eq

\noindent where $n_f$ is the largest principal quantum number
associated with a state at which annihilation is still efficient.
Since the monopole and anti-monopole only annihilate efficiently
when there is a considerable probability of being on top of each
other and this only happens for $ n < 50$, $n_f << n_i$. Thus the
annihilation mean life is small compared with the cascading time and
can be disregarded in the time scale analysis.

The previous equations can be summarized in terms of the binding
energy of the initial bound state and the mean life as,

\be
E_b (eV) \,  r_b (\mbox{\AA}) & \sim & 5. \;  10^4 eV \mbox{\AA} \\
n_i &\sim & 9. \; 10^4 [E_b(eV)^{1/4} \tau(sec)^{1/4}] \\
m c^2 (eV) & \sim & 3. \; 10^7 [E_b(eV)^{3/2}  \tau(sec)^{1/2}]\; eV
\; . \label{binding}\ee

\noindent Here $E_b$ and $r_b$ are respectively the binding energy
and the radius of the initial bound state.

Using Eqs. (\ref{inequality}) through (\ref{binding}) my scenario
can be constructed. I assume that capture takes place for a binding
energy slightly higher, to avoid thermal dissociation, than $kT = 1$
MeV,

\bq E_b > 1 \mbox{ MeV}\; . \eq
This temperature is not related to the scale for production of
monopoles but to that at which monopolium, the bound state, is
formed from already existing monopoles \cite{hill}.

\noindent These equations and the temperature scale of the proposed
scenario show that monopolium is a very tightly bound system

\bq r_b < 0.05 \mbox{\AA} < r_{capture} \sim 0.07 \mbox{\AA}\;
.\label{rcapture}\eq

Let me now proceed to the calculation of the mean free path,

$$ \lambda \sim \frac{1}{\sigma \rho_{ch}}\; .$$
The cross section at $kT = 1$ MeV is given by

$$ \sigma \sim 2.\; 10^{-23}\;\frac{ m c^2(\mbox{eV})}{\mbox{eV}}
\;\mbox{\AA}^2\;. $$
From Eq.(12) I have

$$ \frac{ m c^2(\mbox{eV})}{\mbox{eV}} = 1.5 \; 10^{25}
\;\eta^{\frac{1}{2}}\;, $$
where $\eta = \tau/\tau_U$, $\tau$ being the lifetime of the
monopolium and $\tau_U$ the age of the universe, which we have taken
approximately to be $\tau_U = 1/H_0 \sim 3. \; 10^{17}$ sec.

Thus the cross section becomes

\bq \sigma \sim 3. \; 10^2 \; \eta^{\frac{1}{2}} \; \mbox{\AA }^2
\label{sigma}\eq

I now calculate the density of charged particles. The density of
photons is given by \cite{kolb}

$$ \rho_{\gamma} = \frac{ 2 \zeta (3) (kT)^3}{\pi^2 (\hbar c)^3}\; , $$
while the ratio of the densities of nucleons to photons is given in
nucleosynthesis by $\rho_N/\rho_{\gamma} \sim 4. - 7. \; 10^{-10}$.
At $kT = 1$ MeV , we have mostly nucleons and since $\rho_n/\rho_{p}
\sim \exp{(-1.293/kT)}$, we obtain

$$ \frac{\rho_p}{\rho_{\gamma}} \sim 1. - 3. \; 10^{-10}$$
Due to neutrality for each proton there is an electron, thus

\bq \rho_{ch} \sim 2.5 \;10^{22}\; \mbox{particles/cm}^3.
\label{rhoch}\eq
Thus putting together Eq.(\ref{sigma}) and Eq.(\ref{rhoch}) I obtain

\bq \lambda \sim 0.1  \eta^{\frac{1}{2}} \mbox{\AA}\;. \eq

\begin{figure}[tbp]
\begin{center}
\epsfig{file=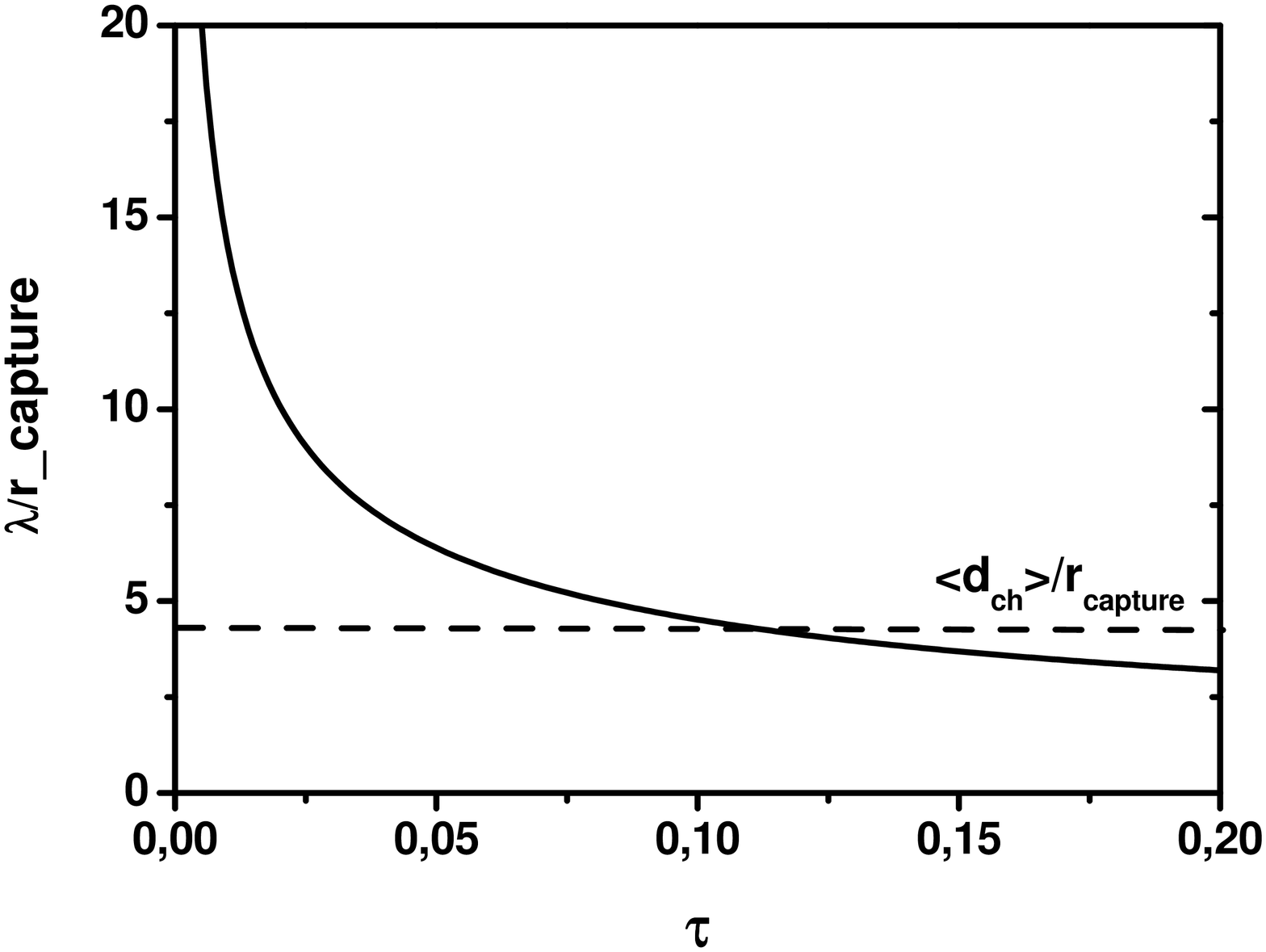,width=8cm,angle=270}
\end{center}
\caption{\small{The ratio $\lambda / r_{capture}$ is presented as a
function of monopolium lifetime $\tau$ in units of age of the
universe $\tau_U$ at $kT = 1$ MeV. Also the ratio of the average
distance between charge particles $<d_{ch}> / r_{capture}$ at the
same temperature is shown. \label{dirac}}}
\end{figure}

In Fig. \ref{dirac} I show the ratio of $\lambda$ over $r_{capture}$
as a function of the life of monopolium in units of the age of the
universe. I also plot the ratio of the average distance between
charges to $r_{capture}$. The plot shows that for lifetimes above
$0.1 \; \tau_U$ the mean free path is smaller than the mean distance
between charges and the capture radius becomes of the order of
magnitude of the mean free path. Thus for monopolium lifetimes of
the order of the age of the universe the conventional scenarios
would take place \cite{hill,sigl,blanco}. However, for $\tau/\tau_U
< 0.1 $ the condition Eq.(2) will be satisfied and a completely
different scenario occurs whose existence has been a matter of
debate since the idea was first proposed \cite{vento}.

In order to clarify the issue let me recall once more the work of
Blanco-Pillado and Olum \cite{blanco}. They pointed out that in the
standard scenario the drag force felt by the monopole in the plasma
reduces dramatically the mean life of the state \cite{blanco} and
therefore relic monopolia do not influence present day observations.
This is so if we are to the left of $\tau/\tau_U > 0.1$. However,
their phenomenon is not active in the scenario presented here, i.e.,
when $\tau/\tau_U < 0.1$ for obvious reasons. As shown in
ref.\cite{goldman} the drag force depends on the two limits of an
impact parameter integral,

$$ F_{drag} \sim \mbox{coefficient} \int_{b_{min}}^{b_{max}} db/b\; ,$$
where naturally the force only arises if $b_{min} < b_{max}$. The
small limit has to do with the monopole interaction with the plasma,
in our case $b_{min} \sim \lambda$. The large limit depends on the
interaction of the monopoles inside the monopolium with the charge
particles and can not extend beyond $r_{capture}$, $b_{max} \sim
r_{capture}$. Due to our assumption Eq.(\ref{inequality}), as seen
in Fig. \ref{dirac}, $r_{capture} < \lambda$ in the region of
interest and therefore in our scenario there is no drag force. It is
interesting to note that the solution that Blanco-Pillado and Olum
find to solve the impasse, i.e. monopoles attached by strings
\cite{blanco}, reproduce with a complex dynamics the same scenario
for monopolium I obtain, i.e. non-relativistic monopoles very
closely tight together so that the light charges do not interact
with them and they only loose energy by radiation.

\section{Monopolium formation}

In the previous section I have shown that the proposed scenario
might be realized if $\eta < < 1$ since under these conditions
primordial monopolium might be observed today. I here study the
other ingredient of the scheme, a plausible scenario for the
formation rate of monopolium which gives consistency to the scheme.
The scheme is based on three time (temperature) scales. The first
time scale is the formation scale of monopoles, $t_i$, whose
corresponding temperature is $T_i$. This temperature  is very high,
consistent with the large mass of the monopoles determined in the
previous section. The second time scale ends at the beginning of the
monopolium bound state stability period, i.e. when the temperature
is lower then the minimum binding energy, i.e. $T_f \sim 1\, MeV $.
Finally the third time scale is today, i.e. $T_{today} \sim 2.7 \,
^0 K$.

In this section I will study the formation rate of monopolium from
the early universe to the beginning of the stability period, i.e.
from $T_i$ to $T_f$. In the next section I will study the decay
process to find present day monopolium abundance, i.e. from $T_f$ to
$T_{today}$.

Following Blanco-Pillado and Olum \cite{blanco} one can write the
evolution equation as

\begin{equation}
\frac{d\Gamma}{dT} = -A \gamma ^2 (T)
\left(\frac{1}{T}\right)^{(9/10)}.
\end{equation}
In this equation $\Gamma$ represents the comoving monopolium
density, i.e. $N_{M \bar{M}}/s$, where $N_{M \bar{M}}$ is the
monopolium density and s the entropy density; $\gamma$ is the
monopole comoving density, i.e. $n_M/s$, where $n_M$ is the monopole
density; $T$ is the temperature and $A$ an softly temperature
dependent quantity.

Let $\gamma$ depend on temperature as

\begin{equation}
\gamma = B T^\delta,
\end{equation}
where B is a constant, a possibility contemplated in ref.
\cite{preskill}. The above equation can be easily solved leading to

\begin{equation}
\Gamma (T_i) - \Gamma (T_f) = - \frac{A}{2 \delta + 0.1}
\left(\gamma ^2 (T_i) \, T_i^{0.1} - \gamma ^2 (T_f)\, T_f^{0.1}
\right)
\end{equation}
My fundamental hypothesis implies,

\begin{eqnarray}
\Gamma (T_i) & << & \Gamma (T_f), \nonumber \\
\gamma (T_i) & >> & \gamma (T_f). \nonumber
\end{eqnarray}
Therefore,

\begin{equation}
\Gamma (T_f) \approx \frac{A}{2 \delta + 0.1} \gamma ^2 (T_i)\,
T_i^{0.1}.
\end{equation}
Dividing by $\gamma (T_f)$ I arrive at,

\begin{equation}
\frac{N_{M \bar{M}} (T_f)}{n_M (T_f)} \approx  \frac{A \;B}{2 \delta
+ 0.1} T_i^{0.1} \left(\frac{T_i^2}{T_f}\right)^{\delta}.
\end{equation}
Thus if $\delta > 0$, since $T_i >> T_f$, the wishful scenario is
realized. Note that this behavior corresponds to $p > 1$ in ref.
\cite{preskill}.

\section{Monopolium abundance}

I proceed in here to investigate phenomenological consistencies of
the proposed picture. Let us calculate the present abundance of
monopolium taking as input the abundance during the formation
period. By doing so my aim is to find consistency between the
proposed scenario and the observation.

The equation governing the density of monopolia $\rho$ taking into
account the expansion of the universe is given by

\bq \dot{\rho} (t) = -\frac{1}{\tau} \;\rho (t) - 3 \;
\frac{\dot{R}}{R} \; \rho (t)\;, \label{rho} \eq
where R is the scale factor of the universe and $\dot{A}$ denotes $d
A/ dt$. If the expansion is adiabatic ($R T \sim$ constant) and the
universe is radiation dominated, the expansion rate is given by

\bq \frac{\dot{R}}{ R} = -  \frac{\dot{T}}{ T} = \frac{
T^2}{\Lambda}\;. \label{expansion} \eq
Here, T is the temperature of the universe and $\Lambda$ a quantity
related to the Planck mass and the effective degrees of freedom
\cite{preskill}. The temperature equation in Eq.(\ref{expansion})
can be integrated to give

\bq t= \frac{\Lambda}{2 T^2} \label{t}\eq
establishing the relation between evolution time and temperature.
Using Eqs.(\ref{rho}), (\ref{expansion}) and (\ref{t}) one obtains,

\bq \dot{\rho} (t) = -\left(\frac{1}{\tau} + \frac{3}{2} \frac{1}{t}
\right) \rho (t) \eq
which can be easily integrated giving

\bq \rho(t) = \rho(t_0) \left(\frac{t_0}{t}\right)^{3/2}
\exp{\left(-\frac{t - t_0}{\tau}\right)}\; . \eq
I now take  this equation, which is  the conventional equation for
the decay of un unstable system in an expanding universe
\cite{kolb}, and adapt it to my interests, namely I want to study
the evolution of the number of monopolia between $T_f$ and
$T_{today}$. Using that, in this case, $t = t_{today} = \tau_U
>> t_0 = t(T_f) \sim 1\, \mbox{sec}$, and the
relation between time and temperature, Eq. (\ref{t}), one can write

\bq \rho (T_{today}) = \rho (T_f)
\left(\frac{T_{today}}{T_f}\right)^{3/2}
\exp{\left(-\frac{\tau_U}{\tau}\right)}\;.\eq
In the present situation this equation reduces to

\bq \rho (2.7 ^0 K) = 1.25 \; 10^{-29} \; \rho (kT= 1 \mbox{MeV})
\exp{(- \frac{1}{\eta})} \;,\label{rho1}\eq
where $\eta = \tau/\tau_U$.

The standard scenario for helium synthesis requires that the mass of
the monopole does not dominate the universe when $kT= 1$ MeV, this
implies \cite{preskill},

\[ \rho_{monopole} (kT = 1 \mbox{MeV}) \le \; 1.3 \; 10^{14}
\left(\frac{m c^2 \; (\mbox{eV})}{\mbox{eV}} \right)^{-1} \; \AA^{-
3}\; . \]
I assume that the density of monopoles is equal to the density of
anti-monopoles and since most of them are bound  equal to the
density of monopolia, thus one gets from Eq.(\ref{rho1}) for the
density of monopolia today,

$$ \rho (2.7 ^0 \mbox{K}) \sim  1.6 \; 10^{-15} \left(\frac{m
c^2 \; (\mbox{eV})}{\mbox{eV}} \right)^{-1} \; \AA^{- 3}\;,
$$
which using Eq. (\ref{binding}) leads to

\bq \rho (2.7 ^0 \mbox{K}) \sim 10^{-16}\; \frac{\exp{(-
\frac{1}{\eta})}}{\eta^{3/2}}\; cm^{- 3}\;. \eq
This equation establishes a relation between the parameter
characterizing the various scenarios $\eta$ and the observation.

\begin{figure}[tbp]
\begin{center}
\epsfig{file=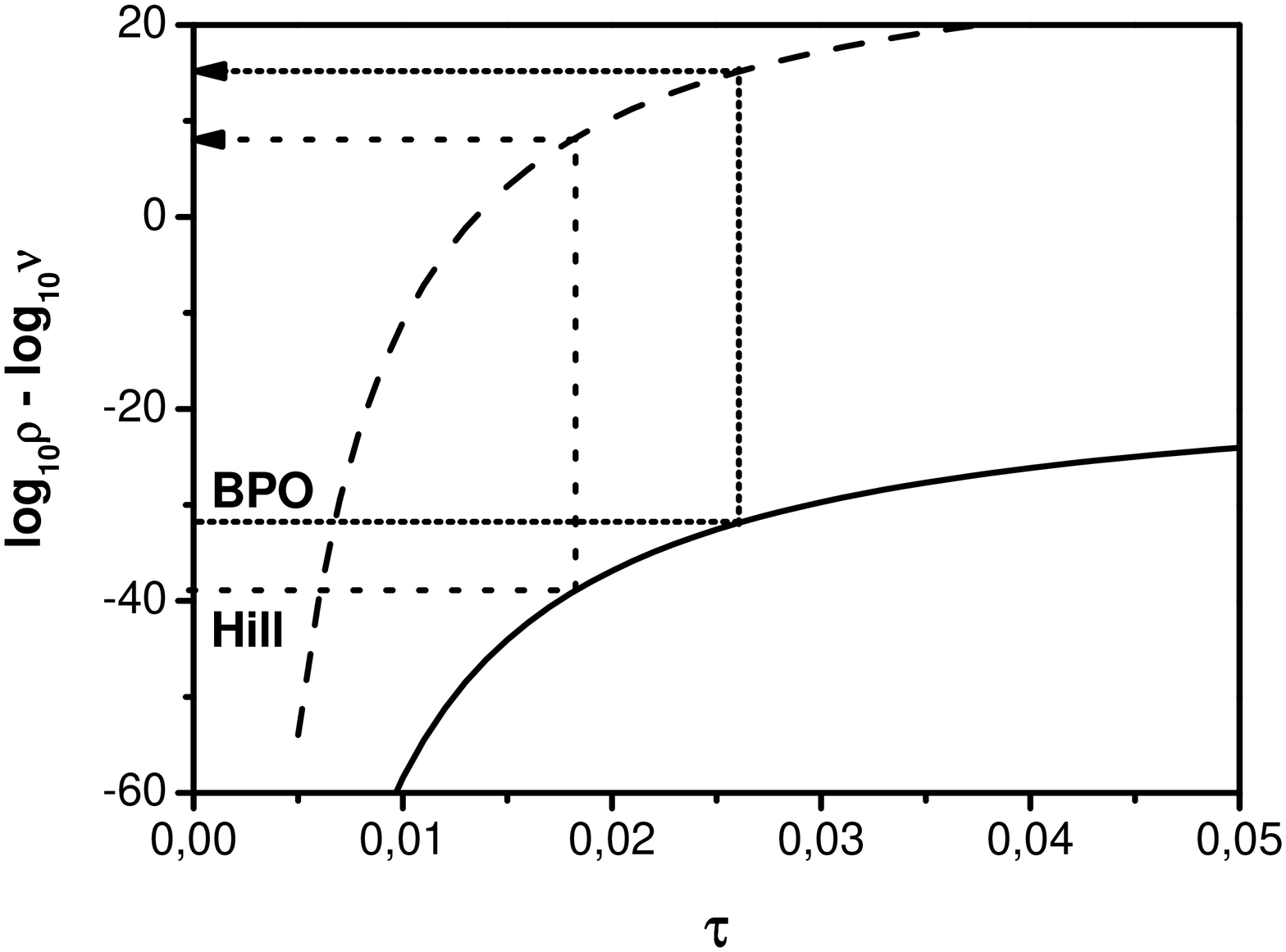,width=8cm,angle=270}
\end{center}
\caption{\small{The density in a $cm^3$ ($log_{10}$) of monopolia
(solid curve) and the number ($log_{10}$)  of monopolia decays in a
$pc^3$ (dashed) today presented as a function of monopolium lifetime
$\tau$ in units of age of the Universe $\tau_U$ . The values by Hill
\cite{hill} (dot) and Blanco-Pillado and Olum \cite{blanco}(short
dot) for the density of monopolia are used to extract the number of
decays.\label{density}}}
\end{figure}

Let me calculate the number of decays per year in a given volume of
the universe

\bq \nu = N \left(1- \exp{\left(- \frac{1}{\tau}\right)}\right) \sim
\frac{N}{\tau (years)} \; ,\eq
where I have used the approximation that $\tau (years) >> 1 \;
year$. Note also that the temperature factor in Eq.(\ref{rho}) drops
out because it is very close to $1$. $N$ is the number of monopolia
in the given volume. Let me choose to calculate the observation a
volume of 1 (pc)$^3$, then I get for the number of events in one
year

\bq \nu \sim 2.7 \; 10^{29} \; \frac{\exp{(-
\frac{1}{\eta})}}{\eta^{3/2}}\; .\eq

Monopolium has been associated with Ultra High Energy Cosmic Rays
(UHECR)\cite{hayashida,bird} in various schemes
\cite{hill,sigl,blanco}. This association leads to a
phenomenological determination of its abundance. I look for
consistency between the phenomenological determined abundances and
the monopolium mean life obtained in the calculation.

\begin{figure}[tbp]
\begin{center}
\epsfig{file=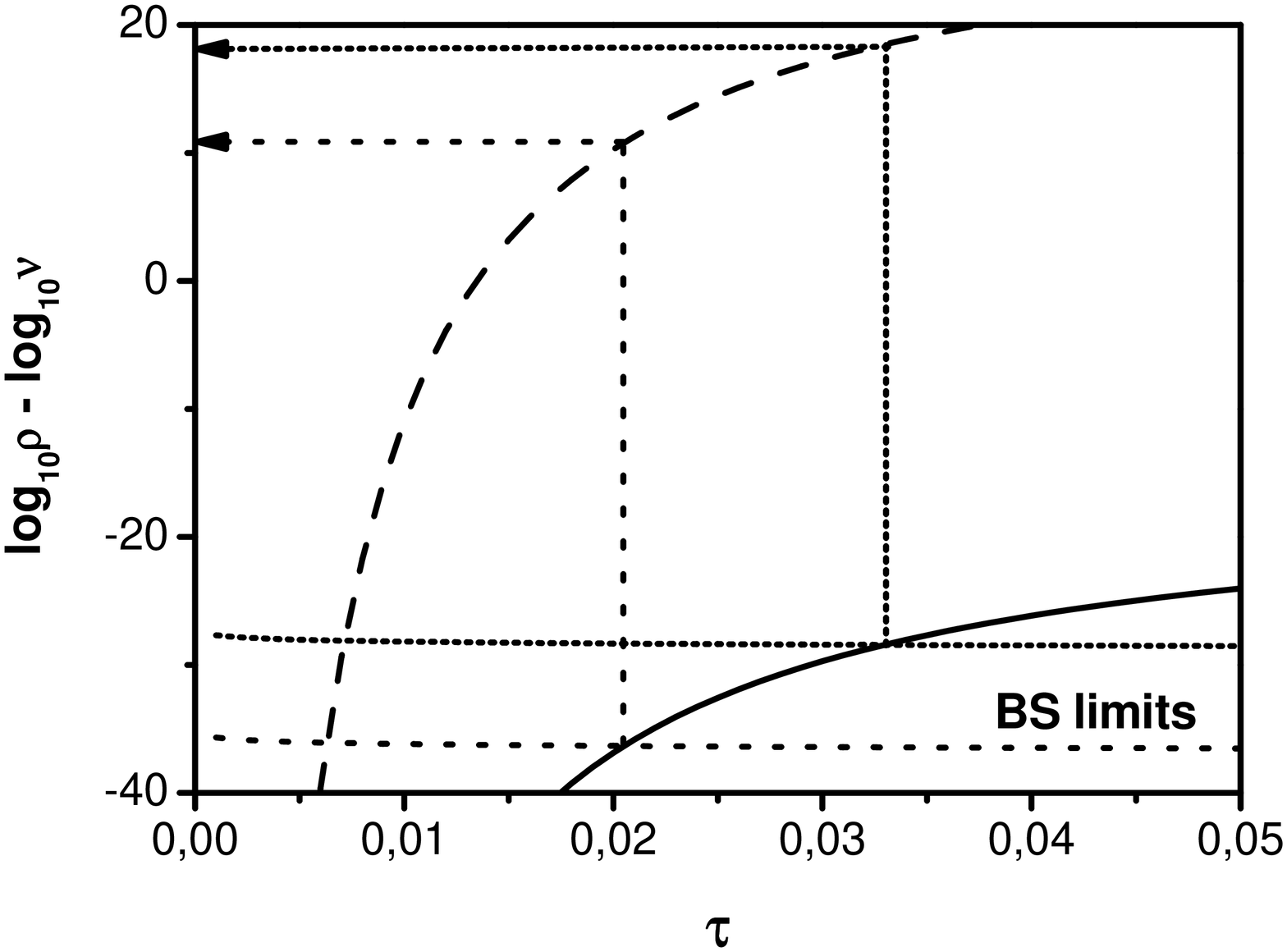,width=8cm,angle=270}
\end{center}
\caption{\small{The density in a $cm^3$ ($log_{10}$) of monopolia
(solid curve) and the number in a $pc^3$ ($log_{10}$)  of monopolia
decays (dashed) today presented as a function of the monopolium
lifetime $\tau$ in units of age of the Universe $\tau_U$ . The
limits of the values  for the density of monopolia from the values
of Bhattarchee and Sigl \cite{sigl}, $1< \Omega_M h^2< 10^8$ are
plotted and the corresponding number of decays
shown.\label{density1}}}
\end{figure}

In Fig. (2) we show the results of our calculation as a function of
$\tau$ in units of $\tau_U$ and I introduce two numerical values for
the density as obtained by

\begin{itemize}

\item[i)] Hill \cite{hill}: $\rho(2.7 ^0 \mbox{K}) \sim 10^{-39} cm^{- 3}$

\item[ii)] Blanco-Pillado and Olum \cite{blanco} :
$\rho(2.7 ^0 \mbox{K}) \sim 10^{-32} cm^{- 3}$

\end{itemize}

In Fig.(3) I compare my results to bands, characterized by $1<
\Omega_M h^2 < 10^8$, obtained from the density equation of
Bhattacharjee and Sigl \cite{sigl}.

The above study shows that proposed values for the density of
monopolia are consistent with $\eta \sim 0.02$ and therefore with

$$\lambda \sim 10 \; r_{capture}$$

If we look at monopolia decays we realize that their number is

$$ \ 10^6 < \nu < 10^{18} \; \mbox{decays/pc}^3 \;$$
which is large compared with those obtained in other models.

Please note that all the estimates are based on the assumption that
all UHECRs are due to monopolia. If they are additional mechanisms
for UHECRs formation or if the number of observations
\cite{hayashida,bird} diminish, the density of monopolia today would
decrease and the proposed scenario would become more natural.
However, this would also imply that it would be harder to confirm
experimentally.

I conclude from the above analysis that at present most monopoles in
the intergalactic vacuum are bound in deep bound states $n \sim 50$
close to the annihilation levels and therefore their binding energy
is at the level of

\bq E_b \sim 10^{14} \mbox{ GeV }  \; ,\eq

\noindent supporting Dirac's conjecture for the non observability of
monopoles.

Moreover, by looking back at Eq.(12) one sees that

\bq m c^2 \sim 2. \; 10^{15}\; \mbox{GeV},\eq
which is very large and comparable to the values arising from GUT
models.

\section{Monopolium detection}

Ideally we would like to be able to produce the monopolium in our
laboratory. Its mass $M\; c^2 \sim 10^{15} \;$ GeV makes laboratory
production impossible. Could we capture a monopolium in our
laboratory and measure its properties? It soon will become clear
this is an impossible task. Monopolium is a sterile particle under
laboratory conditions.

The present day background monopolium density is small.

$$ \rho (2.7 ^0 \mbox{K}) < 10^{-32} \; cm^{-3}. $$
At present monopolia are mostly thermal and therefore their velocity
is of the order

$$ v \sim 10^{-6}\; m/sec .$$
Thus the average number of particles in our detector would be
$$N_M \sim \rho\; v \; t \; A < 10^{- 19} \; (t/years)\; (A/Km^2),
$$
where $A$ is the area of the detector and $t$ the time of exposure.
Thus the chances of having one localized event are insignificant.

Let me assume that a monopolium enters by chance a detector, what
could we observe? Due to the dual behavior of the Maxwell equations
in the presence of monopoles \cite{jackson}, monopolium has an
electric dipole moment

$$ \vec{p} \sim \frac{g}{2\;m \; c} \vec{L} $$
and a magnetic dipole moment
$$\vec{\mu} \sim 2gc < \vec{r}>_{l\ne 0} $$
In the presence of an electric or magnetic field they tend to orient
against the field and become effectively

$$ p\sim \frac{g}{2\;m\;c} m_l $$
and

$$\mu \sim 2\; \frac{\hbar^2\; c}{m\; g}\; (3n^2 - l \; (l + 1)),
 $$
where $n, l, m_l$ are respectively the principal, the orbital and
the magnetic quantum numbers. Substituting the values of the
couplings and masses I get

$$ p \sim 10^{-29} m_l \, \mbox{(e-charge) meter} $$
and
$$ \mu \sim 10^{-22} \; (3 n^2 - l\;(l + 1)) \;\mbox{eV/Tesla} $$
Note that, the maximum achievable electric and magnetic field
gradients today , $50$ MV/meter$^2$ and $5$ Tesla/meter, produce in
background monopolia, with $n \sim l \sim m_l \sim 100$,
insignificantly small energies and forces. It is easy to calculate,
that to stop them, moving at thermal velocities, we would need
distances and times of universe scales.

Thus I conclude, that from the point of view of traditional
laboratory experiments, monopolia are sterile and we have to center
our attention in astrophysical observations.

Let me now turn to astrophysical observations. Let me distinguish
two periods: the formation period and the immediate past.

During the formation period $n \sim 10^9$. Cascading for large
values of $n$ leads to Larmor type emission

\bq \lambda  \sim 16 \alpha ^2 \frac{h}{m c} n_n^3 \sim 32
\frac{\mbox{eV}}{m c^2} n^3\; \mbox{\AA} \; ,\eq
thus the wave length during the nucleosynthesis period will be

$$ \lambda \sim \mbox{milimeter}. $$
Therefore, there should be an isotropic background radio frequencies
as a remnant of that period.

In the immediate past even until today, most monopolia are close to
the annihilating stage, i.e., $n < 100$. The Larmor formula is still
approximately applicable, thus

$$ \lambda < 10^{-15} \; \AA,$$
which implies that the emitted photons will have a huge energy.

$$  h \nu \sim 10^{10}\; \mbox{GeV} $$
This energetic photons will occur at the level of $ 10^{15}-
10^{20}$ per pc$^3$ and therefore should be seen. However, their
distribution will not be isotropic. The core of galaxies, and of
clusters of galaxies, provide an environment of high electric and
magnetic field gradients, thus the small electric and magnetic
dipoles of monopolium will change their distribution in these
environments and create geographic anisotropies.

Moreover, these regions also provide an environment with high energy
and high density where monopolia might be excited to the point of
break up. Thus, also the low frequency spectrum will acquire
geographical distributions which become anisotropies in the spectrum
over the isotropic spectrum remnant of the formation era. Note that
very few monopolia can be formed after electron-positron
annihilation due to the lack of remnant monopoles and the absence of
monopole pair formation due to their huge masses, except in these
very energetic environments, and this phenomenon will not affect
greatly the calculated density of monopolia \cite{blanco}.

Finally, monopolium can annihilate into UHECRs, at the level of
millions per year and per cubic parsec, under present experimental
expectations \cite{hayashida,bird} provided it is the only
mechanism, depositing a huge amount of energy, $E
> 10^{15}$ GeV, in a small region of space-time leading to what
Hill \cite{hill} calls a cataclysmic scenario, whose
details depend on the microscopic theory of monopole formation.

\section{Conclusions}

The possibility of having monopoles in nature is appealing. I have
presented a scenario for the universe in which {\em relic monopoles}
still exist today however, not as free particles, but deeply bound
in monopolium states. The crucial ingredients of my proposal are: i)
that, in the early universe, {\em the mean free path of monopoles is
much larger than their capture radius}, and therefore they bind so
tightly in monopolium that they barely interact with the surrounding
plasma, surviving in this way the effect of the drag force and only
emitting energy by radiation until their annihilation; ii) the
evolution of the monopole density is governed by a strongly
temperature dependent function leading to a large production of
monopolia. Few monopolia are formed in the second period since
almost no free monopoles exist, because they are to produce at these
low temperatures due to their large mass and strong binding within
monopolia, to drive the formation rate. The initial density of
monopolia and their lifetime might explain UHECRs by construction.

Three distinctive quantities determine the consistency of the
various requirements in the scheme: the monopole evolution parameter
$\delta$; the temperature of monopolium formation ($kT \sim 1$ MeV);
and the mean life of the state ($\tau \sim 10^8$ years). The outcome
is monopolium, a neutral particle protected from the interaction
with the medium in a strongly bound state, which radiates copiously
until ultimately annihilates in a cataclysmic scenario being a
possible source UHECRs.

The detection of monopolia, and therefore the existence of
monopoles, presents interesting signatures associated with the
monopolium spectrum, i.e., a diffuse isotropic radio frequency
background remnant of its formation period with geographical
anisotropies produced by more recent activities, and a high
frequency monopolium spectrum associated with the last period of its
lifetime, which manifests itself by anisotropically distributed high
energy gamma rays.

\section*{Acknowledgments}
I thank T. Sloan for a careful reading of the manuscript.
Discussions with Gabriela Barenboim, Jose Bordes, Carlos
Garc\'{\i}a-Canal, Huner Fanchiotti, Pedro Gonz\'alez, Santiago
Noguera and Arcadi Santamar\'{\i}a are acknowledged. This work was
supported by MCYT-FIS2004-05616-C02-01 and GV-GRUPOS03/094.

\end{document}